\documentclass[twocolumn,showpacs,amsmath,amssymb,floatfix,aps,prl]{revtex4}
\usepackage{graphicx}
\usepackage{dcolumn}
\usepackage{bm}
\begin{document}
\title{
Orientational and induced contributions 
to the depolarized Rayleigh spectra \\
of liquid and supercooled ortho-terphenyl}
\author{
S.~Mossa$^{1,2}$, G.~Ruocco$^{2}$, and M.~Sampoli$^{3}$
}
\affiliation{
$^1$ Center for Polymer studies and Department of Physics, 
Boston University, Boston, Massachusetts 02215   \\         
$^2$ Dipartimento di Fisica, INFM UdR  and INFM Center for
Statistical Mechanics and Complexity, 
Universit\`{a} di Roma ``La Sapienza'', Piazzale Aldo
Moro 2, I-00185, Roma, Italy\\
$^3$ Dipartimento di Energetica and INFM, Universit\`a di Firenze, 
Via Santa Marta 3 , Firenze, I-50139, Italy }
\date{\today}
\begin{abstract}
The depolarized light scattering spectra of the glass forming 
liquid ortho-terphenyl have been calculated in the low frequency 
region using molecular dynamics simulation. 
Realistic system's configurations are 
produced by using a recent flexible molecular model 
and combined with two limiting polarizability
schemes, both of them using the dipole-induced-dipole contributions 
at first and second order. The calculated Raman spectral shape are 
in good agreement with the experimental results in a large temperature
range. The analysis of the different contributions to the Raman spectra
emphasizes that the orientational and the collision-induced 
(translational) terms lie on the same time-scale and are 
of comparable intensity. Moreover, the cross terms are 
always found to be an important contribution to the 
scattering intensity. 
\end{abstract}
\pacs{PACS numbers: 64.70.Pf, 71.15.Pd, 61.25.Em, 61.20.p}
\maketitle
\newpage
\section{INTRODUCTION}
\label{introduction}
Depolarized light scattering (DLS) spectroscopy has been proved to be a
valuable tool to explore the dynamical properties of molecular
fluids~\cite{Pecora76}. 
In the case of supercooled and glass forming liquids, DLS has been
used by several researchers with the aim to elucidate 
the dynamical mechanisms underlying the liquid-glass 
transition~\cite{Tao91,Li92,Du94,Wuttke94,Pick94,Lebon95,Alba95}. 
However, the uncertainties in the scattering mechanisms giving rise to
the Raman spectra, and the consequent hypotheses 
that have been introduced to interpret the experiments, 
leave open the problem of a reliable analysis.

In some glass forming liquids, DLS spectra have been interpreted as
arising primarily from interaction-induced mechanisms 
--- namely dipole-induced-dipole (DID) --- 
and so they have been related directly to the dynamics of the density
fluctuations and compared with the outcomings of 
mode-coupling theories (MCT)~\cite{Goetze92}. 
Following this line, Cummins {\it et al.}~\cite{Li92}  
have interpreted the spectra of salol as arising entirely 
from the DID mechanism; in a more recent work~\cite{Cummins96}, 
however, the authors assume that orientational contributions 
dominate the DLS spectra of salol at least up to $4000$ GHz, and 
the connection with the density fluctuations 
dynamics has to rely on a strong coupling between rotational and
translational degrees of freedom. 
The situation is similar also for other systems. 
In the case of an other prototype of fragile glass-forming liquid, 
the ortho-terphenyl (OTP), Patkoswski {\it et al.}~\cite{Pecora97} 
state that the low frequency part of the
spectrum, up to about $10$ GHz, is due to collective reorientation, 
while at higher frequency the DID contribution is dominating. 

We think that a careful investigation  of the scattering mechanisms
which contribute to DLS spectra is needed in order 
to clarify what type of information is possible to extract 
from the experimental data, and to ascertain 
the connections with the density fluctuation correlators treated by MCT.

In the present work we investigate the low frequency DLS spectra of
liquid OTP for temperatures above and close to the 
MCT critical temperature $T_c$. 
We use the new realistic intra- and inter-molecular
potential model set up by some of us very
recently~\cite{Mossa00,Mossa01,soundwaves,fast}. 
Various single particle correlation functions have already been studied 
through classical molecular dynamics (MD) simulations, yielding
noticeable results. It is now possible to combine realistic 
dynamical configurations with plausible polarizability models, 
and try to discriminate among the different hypotheses 
on the origin of the different contributions to DLS experimental spectra.

The paper is organized as follows: Sec.~\ref{model} is a brief 
recall of our potential model and of the methods used to produce 
the dynamical configurations. In Sec.~\ref{polarizability} 
we recall the main polarizability models that have been used so 
far to calculate the DLS spectra of simple molecular liquids; 
moreover, we show how they can be used to estimate the DLS spectra for our
realistic flexible model. 
The computational details are treated in Sec.~\ref{computational}, 
while the results and a discussion are presented 
in Sec.~\ref{discussion}. In Sec.~\ref{conclusions} 
we address some conclusions.
\section{THE MOLECULAR MODEL}
\label{model}
In our flexible model~\cite{Mossa00} the OTP molecule 
is constituted by three rigid hexagons (phenyl rings) 
of side $L_{a}=0.139$ nm. 
Two adjacent vertices of the {\it parent} (central) 
ring are bonded to one vertex of the two {\it side} rings by bonds
whose length at equilibrium is $L_{b}=0.15$ nm. 
Each vertex of the hexagons is thought 
to be occupied by a fictious atom of mass $M_{CH}=13$ a.m.u. 
representing a carbon-hydrogen (C-H) pair. 
In the isolated molecule equilibrium configuration, 
the two lateral rings lie in planes that form an angle of about 54$^o$
with respect to the central ring's plane. The three rings of a 
given molecule interact among themselves by an {\it intramolecular}
potential $V_{intra}$. This potential is chosen in such a way to 
preserve the molecule from dissociation, 
to give the correct relative equilibrium positions 
for the three phenyl rings, and to represent at best 
the isolated molecule vibrational spectrum. 
In particular, it is written in the form 
$V_{intra}=\sum_k c_k V_k({\bm r})$, where each term $V_k({\bm r})$ 
controls a particular degree of freedom; 
the actual values of the coupling constants $c_k$ have been 
determined in order to have a realistic isolated molecule vibrational
spectrum. Several internal motions have been taken into account, 
like the stretching along the ring-ring bonds and between 
the side rings, and the rotation of the 
lateral rings along the ring-ring bonds. 

The intermolecular part of the potential, concerning the interactions 
among the rings pertaining to different molecules, is described 
by a site-site pairwise additive Lennard Jones (LJ) $6-12$ potential, 
each site corresponding to one of the six hexagons vertices. 
The details of the intramolecular and intermolecular 
interaction potentials, together with the values of 
the potential parameters, are reported in Ref.~\cite{Mossa00}. 

It is worth noting that previous studies of the 
temperature dependence of the self diffusion constant~\cite{Mossa00} 
and of the structural ($\alpha$) relaxation times~\cite{Mossa00,Mossa01}, 
indicate that this model is capable to quantitatively 
reproduce the dynamical behavior of the real system, 
but the actual simulated thermodynamical point has 
to be shifted by $\approx$ 20 K upward. In the following, 
when we compare the simulation results with the experiments, 
the reported MD temperatures are rescaled by such an amount.   
\section{POLARIZABILITY MODELS}
\label{polarizability}        
In a liquid composed of $N$ optical anisotropic units 
(a unit can be the entire molecule, a group of atoms 
in the molecule, or a single atom inside the molecule), 
characterized by a permanent polarizability tensor ${\bm \alpha}$, 
the classical low frequency DLS spectrum in the dipole approximation 
is proportional --- apart from a trivial frequency factor --- 
to the Fourier transform of the time correlation function 
of the traceless tensorial part ${\bm \Pi}_2$ of the 
collective polarizability ${\bm \Pi}$.  
${\bm \Pi}$ can be approximated by 
the sum of two terms: the first, ${\bm \Pi}^M$, is the 
sum of all the permanent polarizabilities dependent 
on the orientational variables ${\bm Q}_i$ of the units; 
the second, ${\bf \Pi}^I$, 
is due to all the interaction induced contributions, and its 
leading part can be written in term of the dipole propagator 
tensor ${\bm T}^{(2)}(i,j)={\bm \nabla}^2 R_{ij}^{-1}$ 
(where  $R_{ij}$ is the distance between the i-th and j-th units):
\begin{equation}
{\bm \Pi}={\bm \Pi}^M+{\bm \Pi}^I
\label{eq1}
\end{equation}
\begin{equation}
{\bm \Pi}^M = \sum_i\left\{{\bm \alpha}_i({\bm
Q}_i)\right\}=\sum_i(\alpha {\bm 1} 
+\frac{2}{3}\gamma {\bm Q}_i)
\label{eq2}
\end{equation}
\begin{eqnarray}
{\bm \Pi}^I&=&\sum_i{ \sum_j{ {\bm \alpha}_i {\bm T}^{(2)}(i,j) 
{\bm \alpha}_j}} 
\label{did}\\
&+&\sum_i  \sum_j{  
\sum_k {\bm \alpha}_i {\bm T}^{(2)}(i,j){\bm \alpha}_j 
{\bm T}^{(2)}(j,k) {\bm \alpha}_k }\nonumber \\
&+&\ldots \nonumber
\end{eqnarray}
Here ${\bm \Pi}^M$ is written for units with symmetric top symmetry, 
$\alpha$ and $\gamma$ are the isotropic and anisotropic 
polarizabilities, and ${\bm Q}_{\alpha \beta}=(3/2){\bf \hat{u}}_\alpha 
{\bf \hat{u}}_\beta -(1/2){\bm \delta}_{\alpha\beta}$,
where ${\bf \hat{u}}$ is a unit vector along the symmetry axis. 

Usually only the first order DID is considered in Eq.~(\ref{did}), 
the second order DID being sometimes completely disregarded. 
The relative importance of the higher order DID contributions 
with respect to the first one is dependent on the strength 
of the permanent polarizability tensor, and on the minimum approach 
distances of the units~\cite{Birnbaum85}. 
However, it is well known from the study of noble gas systems,
that at short distances, when high order contributions 
become important, also other  multi-pole and quantum 
electron correlation effects have to be considered.
These terms are usually found to be of opposite sign respect to the DID
contribution~\cite{Dacre82,Hunt85}. 
Empirical diatomic induced polarizabilities 
have negative corrections at short distances for all 
noble gases~\cite{Barocchi85}, and only for gaseous mercury 
a small positive correction to the second order DID 
has been used to fit the experimental spectra~\cite{Sampoli95}. 
From these considerations, in the present work we  
discuss only the contributions from the first and second order DID. 

We want to underline that, in the case of OTP, the choice of the unit 
is not so obvious as for monatomic fluids or simple molecular 
fluids such as $N_2$. Thus, the subsequent choices are based 
on the following considerations. Even for simple molecular systems, 
like $CO_2$ or $CS_2$ fluids, we can consider as unit the entire
molecule, each atom, or the peripheral (most polarizable) atoms. 
In the case more units are pertaining to the same molecule, 
these units are at short distance each other, 
so their induced contributions are tentatively treated in a more refined
scheme. The major attempt in this way has been made 
by Applequist~\cite{Applequist72}.
His formalism, introduced to reproduce the molecular 
polarizabilities with transferable atomic polarizabilities, 
is equivalent to take into account the DID at all orders. 
Indeed the distances are so short that the major 
contributions come from higher order DID interactions. 
Irrespective to the relative success of the method, 
various authors~\cite{Olson78,Thole81,Applequist93}
have modified the scheme adding mono-pole polarizability, 
and/or using non isotropic atomic polarizability, and/or 
separating atoms of the same species on the base of chemical bonds. 
Only a few attempts have been made so far to use the 
Applequist formalism to calculate the experimental 
spectra~\cite{Birnbaum85}, and with doubtful improvements. 
Thus, at present we are not confident in using classical 
dipole interactions at short distances. 

So far we have not tried to use a complex polarizability 
scheme for our molecular model, and we have chosen to adopt 
two models that can give a rough estimate of a lower and 
upper limit of the induced interactions. 
In our potential model, three six sites rigid rings 
form the molecule and this suggests to consider as units 
the rings themselves (R scheme), or the six sites 
of each ring (S scheme). We have always neglected 
the effects on the polarizability 
due to the change of one (or two) H-C with a C-C bond. 
In the R scheme, the polarizability tensor is taken equal 
to that of benzene ($C_6 H_6$). In the S scheme, we have 
simply divided by six the polarizability tensor of benzene, 
and we have neglected the DID contributions coming from 
atoms pertaining to the same ring. The induced contributions 
grow up at least with the sixth power of the inverse 
of the distance, so we expect the R scheme to be a lower limit 
to the induced contributions, while the S scheme to be an upper one. 
Indeed, in the S scheme the mean distances are shorter than in R scheme. 
We want to underline that, if in the S scheme the DID interactions 
inside the ring were taken into account, this would have reduced 
the polarizabilities of the six sites. 

Using a notation similar to that of 
Stassen {\it et al.}~\cite{Ladanyi94}, 
we can work out the collective 
polarizability correlation function for the depolarized scattering:
\begin{equation}
C(t)= \frac{1}{10V}\langle{\bm \Pi}_2(t) {\bm :} {\bm \Pi}_2(0)\rangle,
\label{eq4}
\end{equation}
where $V$ is the volume occupied by the $N$ units. 
$C(t)$ can be divided in three terms, namely the 
orientational term $C_{or}(t)$, 
the interaction-induced or DID term $C_{ii}(t)$, 
and the cross contribution term $C_{cr}(t)$:
\begin{equation}
C(t)=C_{or}(t) + C_{ii}(t) + C_{cr}(t),
\end{equation}
with
\begin{eqnarray}
C_{or}(t)&=&\frac{2}{45V} \gamma^2 \langle {\bm Q} (0) {\bm :} 
{\bm Q} (t) \rangle
\nonumber\\
C_{ii}(t)&=& \frac{1}{10V} 
\langle{\bm \Pi}_2^I(0) {\bm :} {\bm \Pi}_2^I(0)\rangle
\label{eq5}\\
C_{cr}(t)&=& \frac{1}{15V} \gamma \left[
\langle{\bm Q} (0) {\bm :} 
{\bm \Pi}_2^I(t)\rangle + \langle{\bm \Pi}_2^I(0) 
{\bm :} {\bm Q} (t)\rangle\right].
\nonumber
\end{eqnarray}
Here ${\bm Q} = \sum_i{{\bm Q}_i}$ and all the sums 
are extended over the $N$ units. 

Although the importance of the three 
terms in Eq.~(\ref{eq5}) has been found 
similar in various systems~\cite{Birnbaum85}, 
the cross term $C_{cr}(t)$ has been neglected, 
without sound arguments, in various works 
on glass-forming liquids. The integrated intensities of 
the orientational contribution, i.e., the value at $t=0$ 
of the correspondent correlation function, 
can be written as:
\begin{equation}
C_{or}(t=0) = \frac{N}{15 V} \gamma^2 (1+f_2),
\label{eq6}
\end{equation}
where $g_2=1+f_2$ is the static angular pair correlation factor 
between our units, defined as~\cite{Gershon69}
\begin{equation}
g_2 \equiv \frac{2}{3N} \sum_i \sum_{j \neq i}\langle{\bm Q}_i (0) 
{\bm :} {\bm Q}_j (0)\rangle.
\label{eq6a}
\end{equation}
It is well known that, if the units are spherically
symmetric ($\gamma=0$), the orientational and cross 
contributions do vanish, as in the case of
DLS from monatomic fluids. 
In both R and S schemes, we expect a value of $g_2$ rather different
from $1$. Indeed, also in the R scheme we have a 
strong orientational correlation between different units 
(rings pertaining to the same molecule). 
This is quite different from the case of simple molecular fluids, such
as $N_2$, where only very small orientational correlations exist between
different molecules, even in the liquid phase near the melting point. If
all the rings of OTP molecules were in the equilibrium positions, while
there was no correlation between the orientation of different molecules,
we can calculate the $g_2$ factor for both R and S models. Using the
equilibrium values of the cosines between the normals to the ring
planes (${\bf n}_1 \cdot {\bf n}_2 = {\bf n}_1 \cdot {\bf n}_3=0.588$,
and ${\bf n}_2 \cdot {\bf n}_3 =0.673$, 
with the index $1$ standing for the 
central ring~\cite{Mossa00}), we have
\begin{equation}
g_2^{R, eq} \approx 1.14,\;\;g_2^{S, eq} 
= 6 \times g_2^{R,eq} \approx 6.84.
\label{eq9}
\end{equation}
Obviously, the obtained $g_2$ factors make the orientational 
contribution integrated intensity --- in this idealized 
condition --- independent of the adopted scheme. 
Since the relation $g_2^{R, eq}=g_2^{S, eq}/6$ holds always, 
in the following we refer simply to $g_2$ in the R scheme. 
Further, we can evaluate the average value of the analogous 
of $g_2$ for the rings pertaining to the same molecules, i.e.
\begin{equation}
K_2 = \frac{3}{2N} \sum_i \sum_{j \neq i} 
\langle \cos^2(\theta_{ij})-1/3 \rangle,
\label{eq6b}
\end{equation}
where the sum is extended over the rings of a molecule, 
$\theta_{ij}$ is the angle between the normals 
of the i-th and the j-th ring, and $\langle \rangle$ 
means a configuration or time average. 
The value of $K_2$ for the isolated molecule at equilibrium 
is always $K_{2}^{eq} \simeq 1.14$.
\begin{figure}[t]
\centering
\includegraphics[width=.4\textwidth]{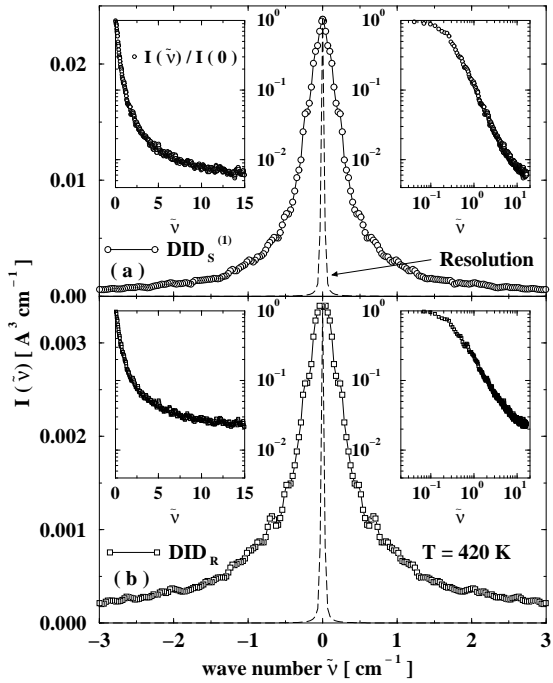}
\vspace{-0.5cm}
\caption{(a) First order DID spectrum 
calculated considering the carbon atoms (model S) 
at $T=420$ K together with the resolution curve (see text for
details). The insets show the data, renormalized to the
peak intensity,  in semi-logarithmic (left
panel) and logarithmic scale (right panel), 
in order to stress the power law behavior of the tail. 
(b) DID spectrum calculated considering the phenyl rings 
centers of mass (model R).}
\label{fig1}
\end{figure}
\begin{figure}[t]
\centering
\includegraphics[width=.4\textwidth]{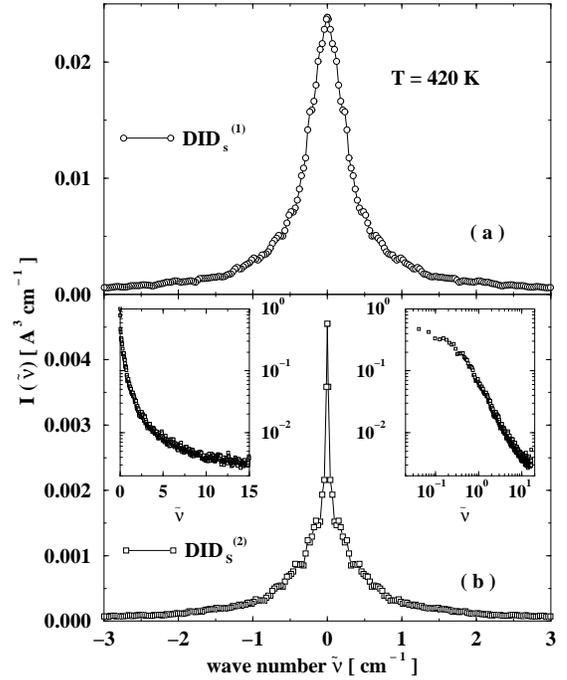}
\vspace{-0.5cm}
\caption{(a) As in Fig.~1(a). 
(b) Second order DID calculated for model S.}
\label{fig2}
\end{figure}
\begin{figure}[b]
\centering
\includegraphics[width=.4\textwidth]{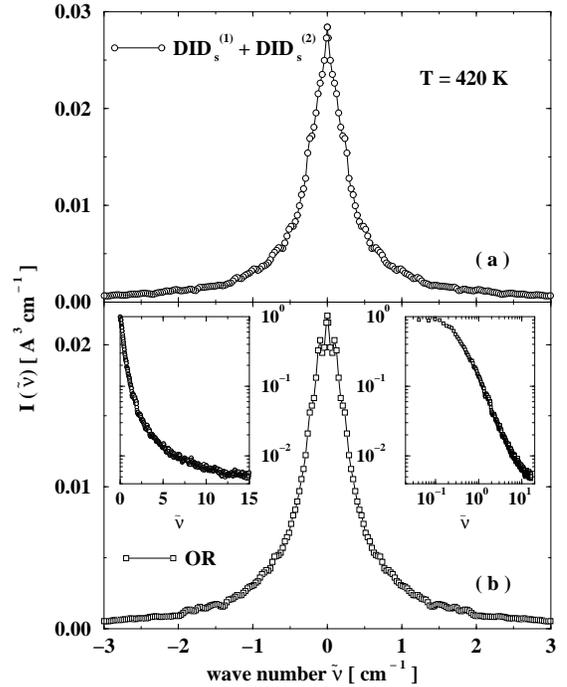}
\vspace{-0.5cm}
\caption{(a) Sum of first and second order 
DID terms for the $S$ model. (b) Orientational 
contribution to the spectrum.}
\label{fig3}
\end{figure}
\begin{figure}[tb]
\centering
\includegraphics[width=.4\textwidth]{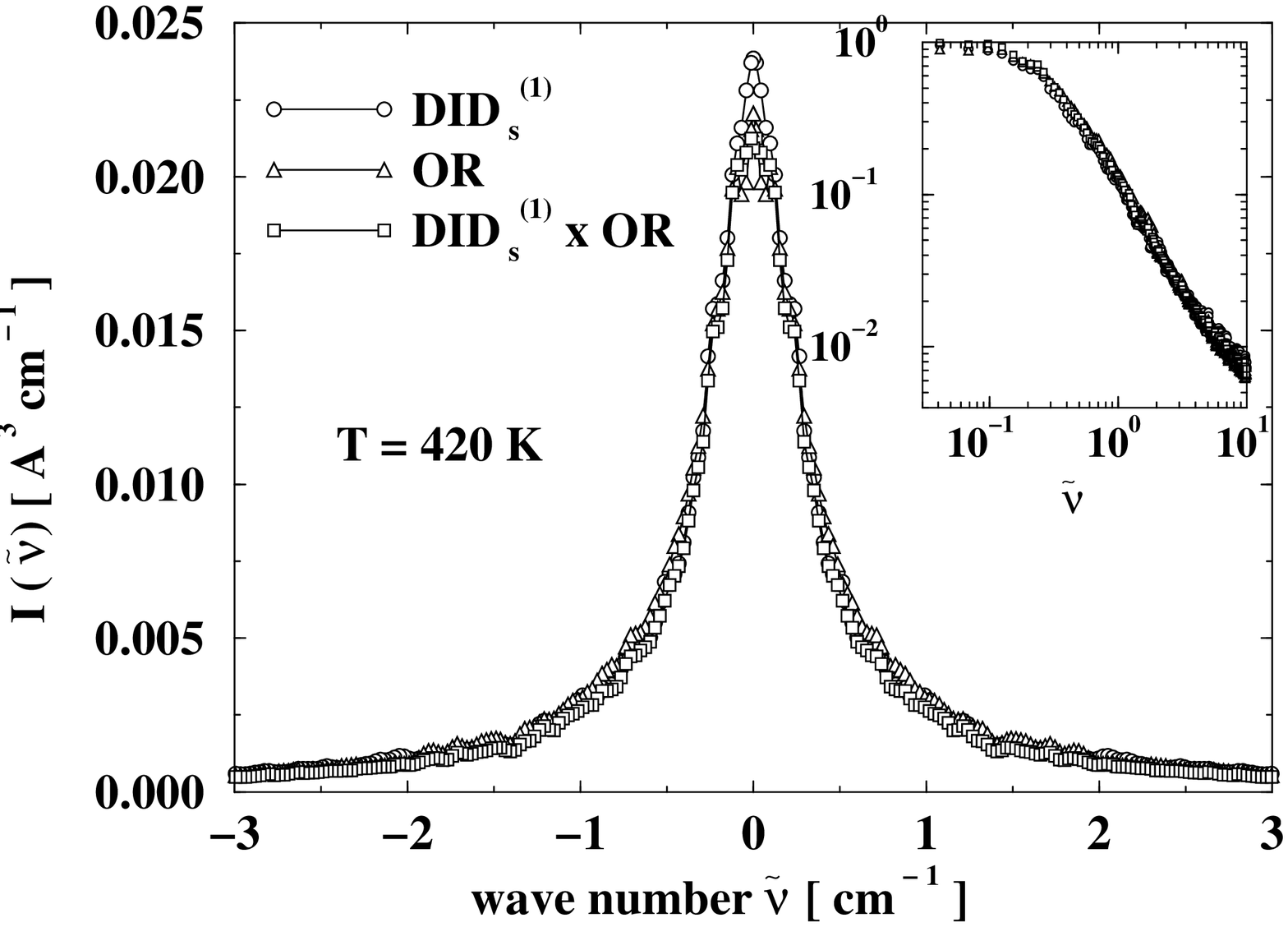}
\caption{Contributions to the total 
spectrum for the S model: first order DID
(circles), orientational contribution (triangles),
and cross-correlation (squares) calculated at $T=420$ K; 
the inset shows the same data, renormalized to the
respective peak intensities, in a double-logarithmic scale.}
\label{fig4}
\end{figure}
\begin{figure}[tb]
\centering
\includegraphics[width=.4\textwidth]{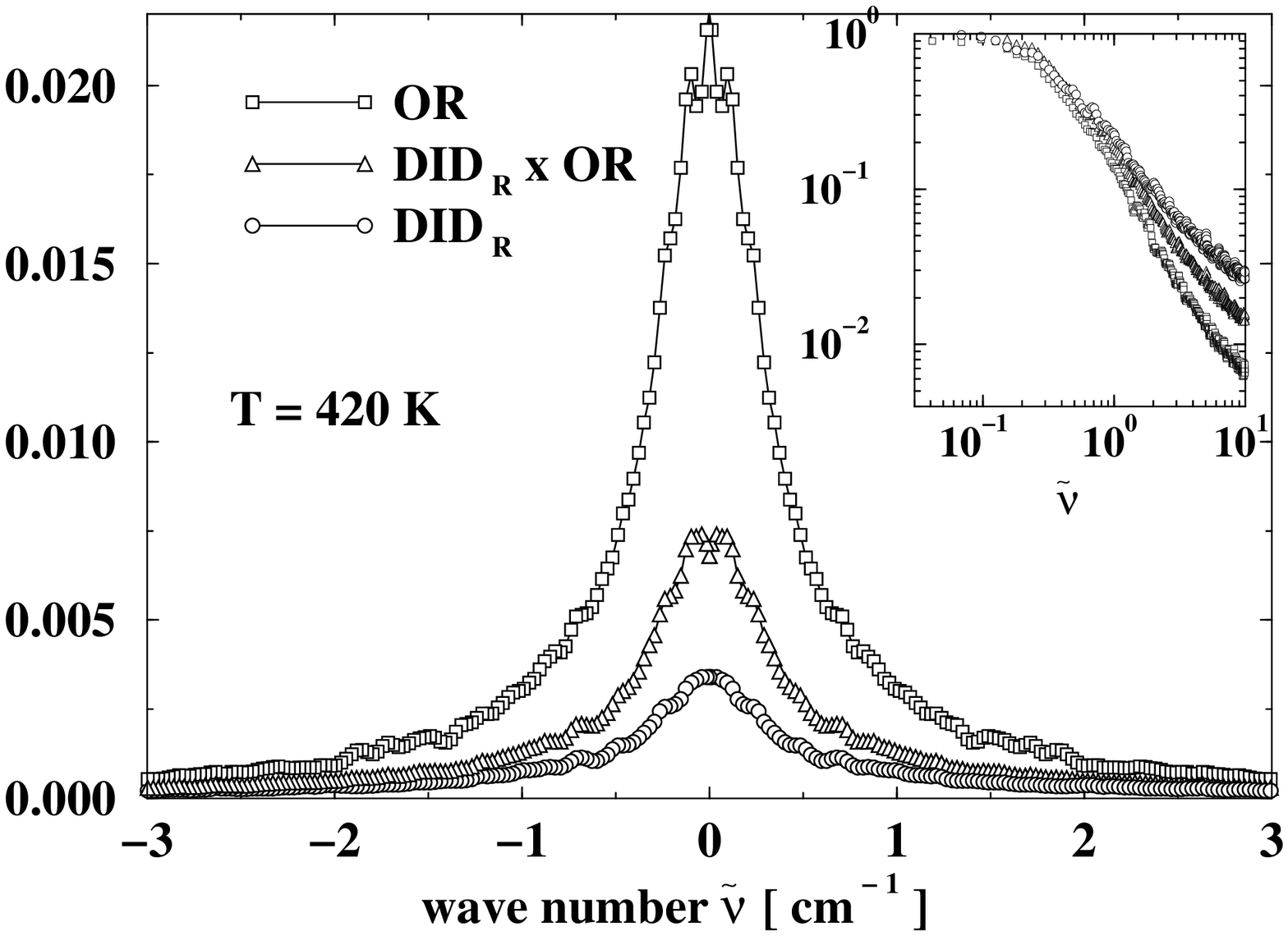}
\caption{Contributions to the total spectrum for the $R$
model: first order DID (circles), orientational contribution (triangles) 
and cross-correlation (squares) calculated at $T=420$ K; the inset shows
the same data, renormalized to the respective peak intensities, 
in a double-logarithmic scale.}
\label{fig5}
\end{figure}
\begin{figure}[tb]
\centering
\includegraphics[width=.4\textwidth]{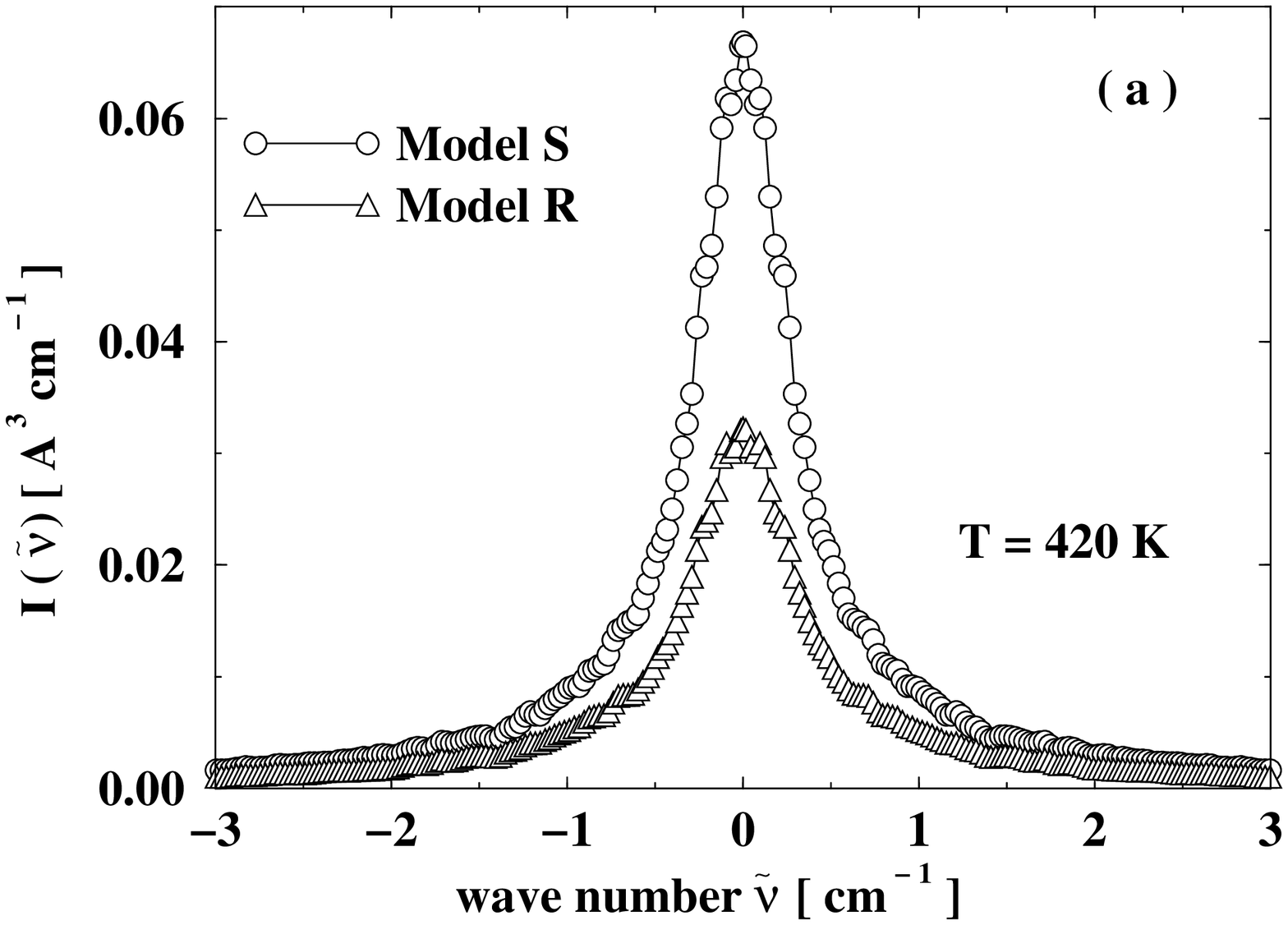}

\includegraphics[width=.4\textwidth]{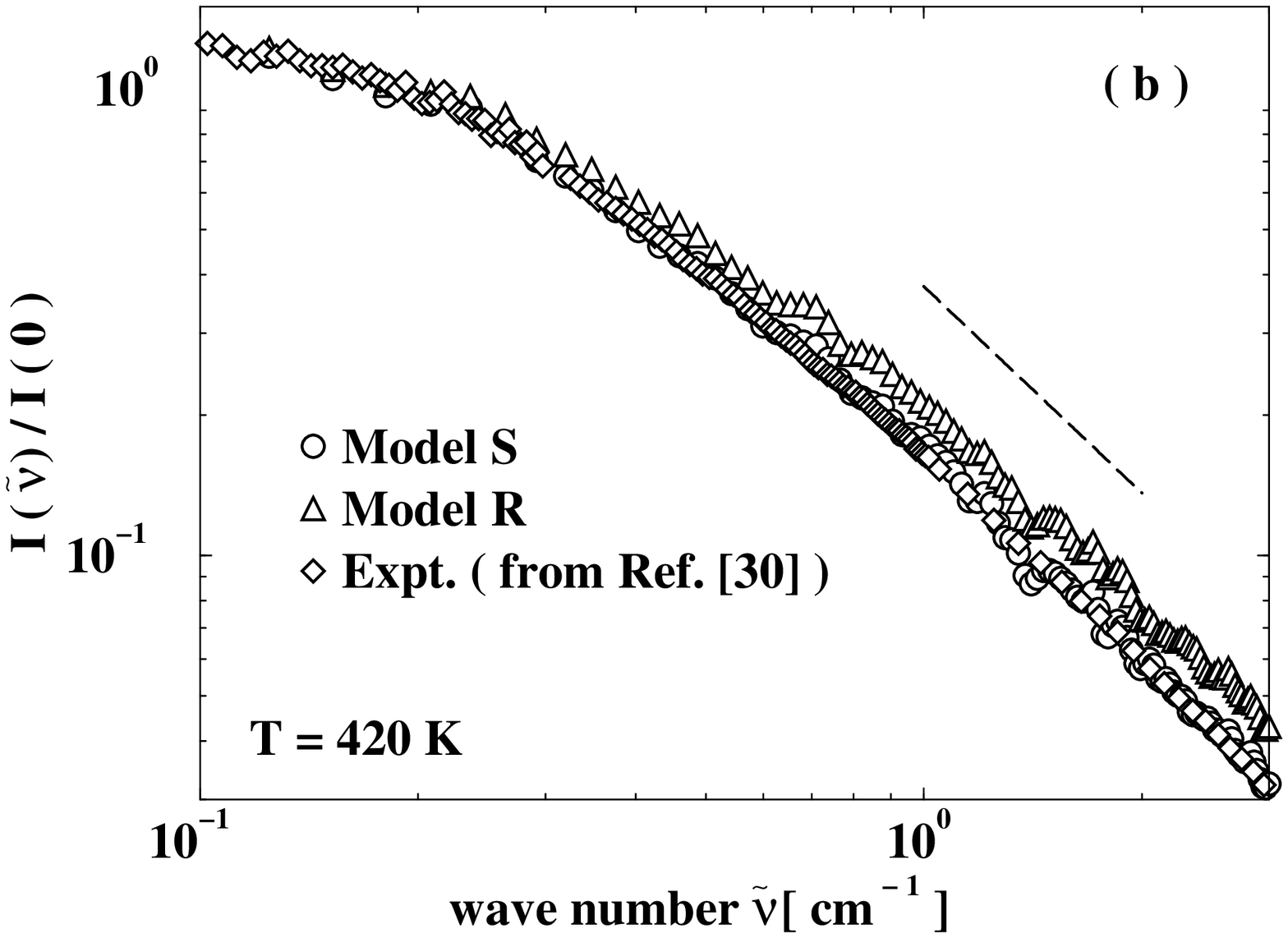}
\caption{(a) Total spectra 
for the S (circles) and R (triangles) models. 
(b) Total spectra for the S (circles) and R (triangles) models
together with the experimental results ($V_H$ spectra
from Ref.~\protect\cite{Angelini99}) at $T=444$ K (diamonds)
in a double logarithmic scale. 
A power law of exponent $\lambda=-1.5$ (dashed line)
is also plotted as a guide to the eye.}
\label{fig6}
\end{figure}
\begin{figure}[tb]
\centering
\includegraphics[width=.4\textwidth]{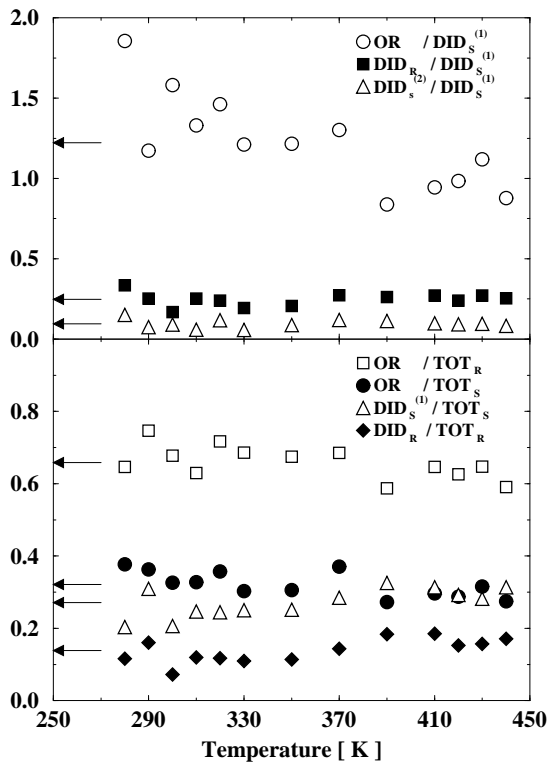}
\caption{(a) Ratios between the first order 
DID contribution to the $S$ model
and the other terms as a function of temperature: Orientational
(circles), DID calculated for model R (full squares) and second order DID
calculated for model S (triangles). The arrows indicate the mean values
that are respectively $1.22$, $0.25$, and $0.09$.
(b) Relative intensity of the different 
contributions with respect to the total
intensity calculated with both models. 
The arrows indicate the mean values of respectively 
$0.66$, $0.32$ $0.27$, and $0.14$.}
\label{fig7}
\end{figure}
\begin{figure}[tb]
\centering
\includegraphics[width=.4\textwidth]{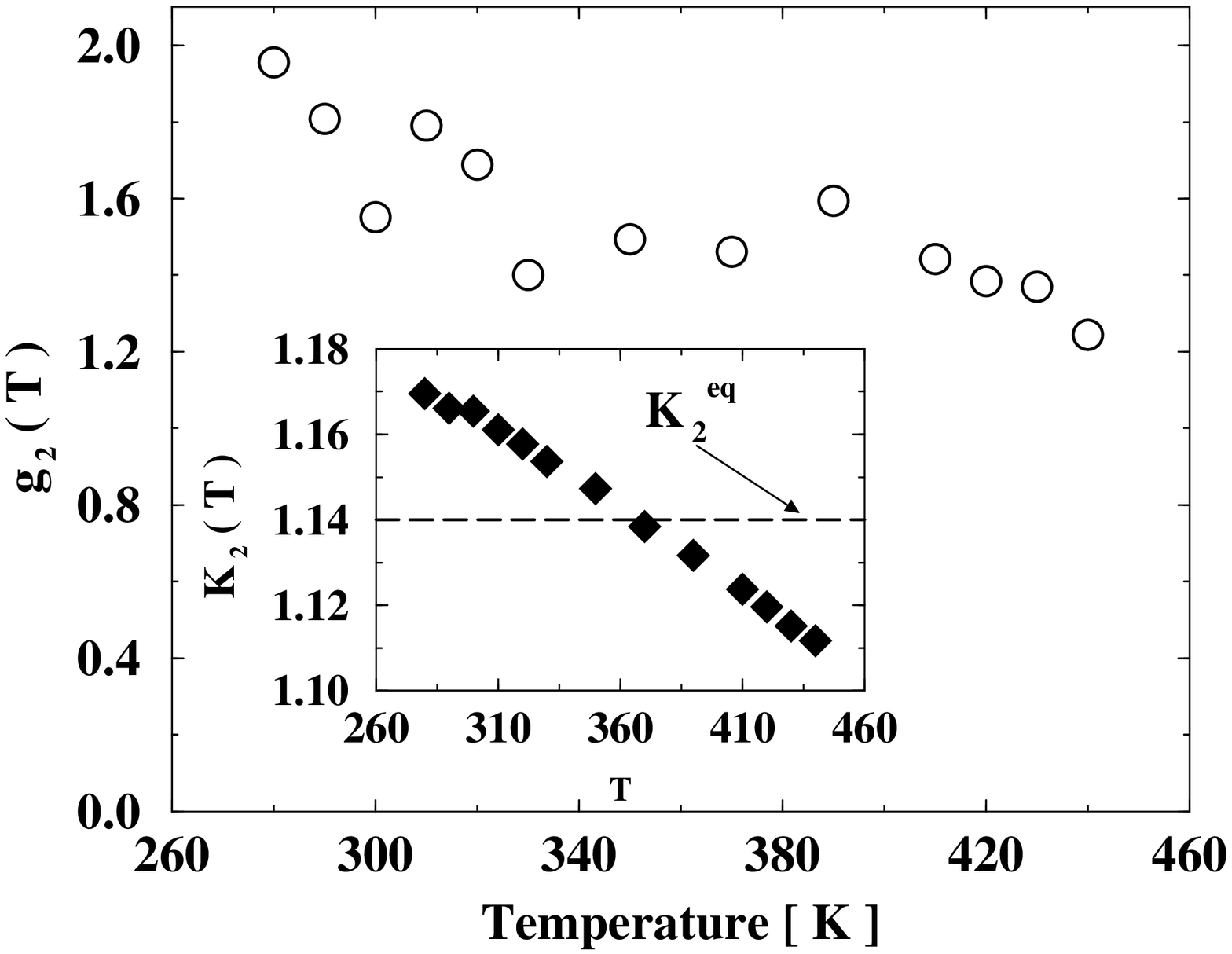}
\caption{Static angular pair correlation 
factor $g_{2}(T)$ (Eq.~(8)) as a function of temperature. 
In the inset, $K_{2}(T)$ (Eq.~(10)) is shown as 
a function of temperature.}
\label{fig8}
\end{figure}
\section{COMPUTATIONAL DETAILS}
\label{computational}
We study a system composed by $108$ OTP molecules 
($324$ rings, $1944$ LJ interaction sites)
enclosed in a cubic box with periodic boundary condition.
The MD runs are performed in the  microcanonical (NVE) ensemble.
To integrate the equations of motion we treats
each ring as a separate rigid body, identified by the position 
of its center of mass ${\bm R}_{i}$ and by its orientation, 
expressed in terms of quaternions ${\bm q}_{i}$~\cite{allen}. 
The standard Verlet leap-frog algorithm~\cite{allen} 
has been implemented to integrate the translational motion while, 
for the most difficult orientational part, 
a refined algorithm has been used~\cite{ruosam}. 
The integration time-step is $\Delta t$=2 fs, 
which gives rise to an overall energy conservation better 
than 0.01 \% of the kinetic energy. 

We consider two series of runs. For temperature $T=420$ K, 
to be compared with the experimental results, we perform 
one run $24$ ns long with a saving time of $1$ ps; 
using windows $800$ ps wide we achieve a resolution of 0.04 cm$^{-1}$.
For the temperatures $T=$ 280, 290, 300, 310, 
320, 330, 350, 370, 390, 410 and 430 K 
we run for $640$ ps with a saving time of $0.04$ ps; 
using windows of $320$ ps we reach a resolution of 0.1 cm$^{-1}$.

In the R scheme, the components of the polarizability tensor
(in the molecular fixed frame) are
$\alpha_\parallel=12.31$ $\AA^3$, $\alpha_\perp=6.35$ $\AA^3$, 
$\alpha=1/3(\alpha_\perp+2\alpha_\parallel)=10.32$ $\AA^3$,
$\gamma=\alpha_\perp-\alpha_\parallel=-5.96$ $\AA^3$,
where $\perp$ stands for orthogonal to the phenyl ring's plane, 
i.e. parallel to the symmetry axis.
These values are taken from the polarizability tensor 
of the benzene~\cite{hirsch}.
For the S scheme, to each site have been attributed 
the previous values divided by $6$ and the same orientation.

\section{RESULTS AND DISCUSSION}
\label{discussion}
In Fig.~1 the DID spectrum at first order is reported 
in the liquid phase at $420$ K for both S (Fig.~1(a)) 
and R (Fig.~1(b)) schemes. The dashed lines refer to 
the obtained resolution, which depends on the time length of 
the MD run. In the insets the same data, rescaled
to the peak intensity, are shown in semi-logarithmic (left) 
and logarithmic (right) scale in order to emphasize 
the power law behavior of the tails. 

As expected, the overall intensity in the S scheme is much 
higher than the one in the R scheme. On the contrary, the shape 
is very similar, although  the relative intensity of the 
spectral wings (at 10 $cm^{-1}$) with respect to the 
peak intensity is about three times higher in the R scheme. 
As a consequence the absolute wing intensities are comparable. 
On the overall, the shape is similar to the induced 
contribution from heavy noble gases, i.e., an exponential 
decay of the form $\exp(-\omega / \omega_o)$ 
with $\omega_o$ of the order of 2 $cm^{-1}$. 

Only for the S scheme the second order DID is of some importance, 
so in Fig.~2 the corresponding spectrum (Fig~2(b))
is compared with the first order DID (Fig~2(a)). At all frequencies
the second order contribution is a relatively small fraction 
of the first one; above 0.25 cm$^{-1}$ it is less 
than about 10$\%$, so the DID's at higher order can be safely neglected. 

The sum of the first and second order DID, and 
the orientational spectra in the S scheme 
are compared in Fig.~3(a) and (b) respectively. 
As we can see, the relative shapes and  intensities 
in this case are quite similar 
(apart from the shape of  the peak at frequency comparable 
to the spectral resolution, where the simulation 
statistical errors are large and prevent any reliable comparison). 
We want to stress that no time scale separation is 
detectable between translational and orientational dynamics. 
This is reflected also in the shape and high intensity of 
the cross contribution; at this temperature, indeed, 
the spectra of orientational, DID and cross 
contributions practically superimpose each other, 
as shown in Fig.~4.

In the case of the R scheme, the situation is quite different. 
Beyond the smaller relative importance of the DID and cross terms, 
at high frequency (above 5 cm$^{-1}$) the spectrum is  
dominated by the DID contributions, even if at 10 cm$^{-1}$ 
the ratio of DID, cross and orientational contributions 
is not so high, being $4:2:1$ (see Fig.~5).

Again, even in R scheme our simulations are in poor agreement 
with the assumptions made by Patkowski {\it et al.}~\cite{Pecora97}
about the spectral separation between orientational and DID
contributions. Indeed, it is evident from Fig.~5
that no time scale separation can be effective and the 
cross term cannot be neglected. 

The resulting --- total --- simulated DLS spectra for the two schemes 
are reported in Fig.~6(a). In Fig.~6(b) we show, 
in double logarithmic scale, a comparison of the two line shapes 
with the experimental spectrum at $444$ K~\cite{Angelini99}. 
The agreement among the sets of data is quite good, 
considering the approximations we have made in the dynamical 
and polarizability model. 
This shape comparison gives some preference to the S scheme, 
but it is hard to really discriminate between these 
two limiting polarizability models on this basis. 
R and S DLS spectra differ essentially in intensity 
rather then in shape, and some help would come from an 
absolute intensity comparison; unfortunately, the OTP 
experimental absolute intensity has not been calculated
so far. Anyway, the large uncertainties usually present 
in both experimental and simulated determinations 
of DLS absolute spectral intensities, could not allow
this kind of comparison.

In Fig.~7 we plot the temperature dependence 
of the relative integrated intensities of the various spectral
contributions. We see that the relative intensities do not show 
any significant trend with temperature, apart from a tendency 
of the orientational part to increase at the lowest 
investigated temperatures. The increase has to be attributed 
to a change in the relative orientation of the rings, i.e., 
to an increasing of $g_2$. This is clearly seen in Fig.~8
where $g_2$ (main panel) and $K_2$ Inset) 
are plotted as a function of temperature. 
It is interesting to note that also $K_2$ increases on 
decreasing temperature, being less than the reference 
value $K_{2}^{eq}\simeq$ 1.14 at high temperatures and 
more than that value at low temperatures. 
On the basis of the value of the angles between 
the symmetric top axis of the rings, $K_2$ increases 
fast if there is a spread of ${\bf n}_1 \cdot {\bf n}_2$ 
and ${\bf n}_1 \cdot {\bf n}_3$, but decreases with 
the spread of ${\bf n}_2 \cdot {\bf n}_3$.
The distribution of angle cosines is reported in Fig.~6 of 
Ref.~\cite{Mossa00}. At high temperatures, the large spread 
of ${\bf n}_2 \cdot {\bf n}_3$ is, anyway, able to decrease  
$K_2$ under $K_{2}^{eq}$, but at low temperature the 
situation is reversed by the spread of ${\bf n}_1 \cdot 
{\bf n}_2$ and ${\bf n}_1 \cdot {\bf n}_3$.

We expect a similar situation to hold for the DLS of a 
large class of molecular glass forming liquids:
the DLS would be a mixture of orientational, 
DID and cross contributions in the entire low frequency range, 
with no significant time scale separation. 
Further, the better agreement with the experimental results 
of the S scheme in the case of OTP, 
underlines the importance to take into account 
the internal degrees of freedom to obtain a realistic 
description of the DLS of glass forming 
liquids consisting of large flexible molecules.
\section{CONCLUSIONS}
\label{conclusions}
In this paper we have studied, by means of molecular dynamics 
simulations, the orientational and induced contributions 
to the low frequency depolarized Rayleigh spectra 
of supercooled ortho-terphenyl. We have used a 
realistic flexible intramolecular model recently introduced 
by some of us, taking into account the most important 
internal degrees of freedom of the OTP molecule. 
Two polarizability models have been
introduced, each of them considering a different scattering unit. 
In the ring's model R, we have assigned to each phenyl 
ring the polarizability of the benzene. In the 
site's model S, we have assigned to the sites of each 
ring the same polarizability divided by six. 
This procedure has allowed us to study the single contributions 
to the DLS spectra in both cases. Our main findings are the following:
{\it i)} Although in the two schemes the intensities of the DID 
contributions are very different, their overall shape is very similar;
{\it ii)} The second order DID is at all frequencies already 
a relatively small fraction of the first order contributions,
so that all the higher induced terms can be safely neglected;
{\it iii)} For the S model first order DID and orientational 
contributions are very similar in both shape and intensity;
{\it iv)} In both models the cross correlation between 
induced and orientational terms cannot be neglected. 
This fact is in striking contrast with the analysis of 
the experimental data of Ref.~\cite{Pecora97}
and support the conclusion that: 
{\it v)} No time scale separation is present between DID 
and orientational contribution. In other words we expect that, 
similarly to the present case, for a broad class of molecular 
glass forming liquids the DLS spectra would be a superposition of DID, 
orientational and mixed contributions,
and none of them should be disregarded. Finally: 
{\it vi)} the better agreement of the S scheme calculated spectra 
with the experimental results underlines the importance 
of taking into account the molecular internal degrees of 
freedom in order to obtain a realistic description 
of complex molecular liquids. 

\begin{center}
{\bf ACKNOWLEDGEMENTS}
\end{center}

This work was supported by MURST PRIN 2000.
%
%
%
%

%
%

%
%
\end{document}